\documentclass[10pt,twocolumn]{IEEEtran}
\usepackage{graphicx}
\usepackage{amsmath}
\usepackage{amssymb}
\usepackage{mathrsfs}
\usepackage{stfloats}
\usepackage{dsfont}
\usepackage{setspace}
\usepackage{array}
\usepackage{bbm}
\usepackage{mathabx}
\usepackage{algorithmic}
\usepackage{algorithm}
\usepackage{glossaries}

\newcommand{\eeq}{\end{equation}}

\newcommand{\bh}{\mbox{\boldmath $h$}}

\newcommand{\bPhi}{\mbox{\boldmath $\Phi$}}

\newcommand{\bg}{\mbox{\boldmath $g$}}

\newcommand{\ds}{\displaystyle}

\newcommand{\beq}{\begin{equation}}

\newtheorem{proposition}{Proposition}

\makeglossaries

\newacronym{mac}{MAC}{multiple-access channel}
\newacronym{bc}{BC}{broadcast channel}
\newacronym{mimo}{MIMO}{multiple-input multiple-output}
\newacronym{siso}{SISO}{single-input single-output}
\newacronym{sc}{SC}{single-carrier}
\newacronym{mc}{MC}{multi-carrier}
\newacronym{ofdma}{OFDMA}{orthogonal frequency division multiple access}
\newacronym{af}{AF}{amplify-and-forward}
\newacronym{df}{DF}{decode-and-forward}
\newacronym{cf}{CF}{compress-and-forward}
\newacronym{mwrc}{MWRC}{multi-way relay channel}
\newacronym{pde}{PDE}{partial data exchange}
\newacronym{fde}{FDE}{full data exchange}
\newacronym{iid}{i.i.d.\@}{independent and identically distributed}
\newacronym{awgn}{AWGN}{additive white Gaussian noise}
\newacronym{awg}{AWG}{additive white Gaussian}
\newacronym{sic}{SIC}{successive interference cancellation}
\newacronym{snr}{SNR}{signal-to-noise ratio}
\newacronym{sinr}{SINR}{signal-to-interference-plus-noise ratio}
\newacronym{ber}{BER}{bit error rate}
\newacronym{zf}{ZF}{zero-forcing}
\newacronym{mmse}{MMSE}{minimum mean square error}
\newacronym{sud}{SUD}{single user decoding}
\newacronym{dof}{DoF}{degrees of freedom}
\newacronym{gdof}{GDoF}{generalized degrees of freedom}
\newacronym{nnc}{NNC}{noisy network coding}
\newacronym{dmn}{DMN}{discrete memoryless network}
\newacronym{csi}{CSI}{channel state information}
\newacronym{ee}{EE}{energy efficiency}
\newacronym{ian}{IAN}{treating interference as noise}
\newacronym{snd}{SND}{simultaneous non-unique decoding}
\newacronym{brd}{BRD}{best response dynamics}
\newacronym{br}{BR}{best response}
\newacronym{ne}{NE}{Nash equilibrium}
\newacronym{lhs}{LHS}{left-hand side}
\newacronym{rhs}{RHS}{right-hand side}
\newacronym{gee}{GEE}{global energy efficiency}
\newacronym{wsee}{WSEE}{weighted sum energy efficiency}
\newacronym{wpee}{WPEE}{weighted product energy efficiency}
\newacronym{wmee}{WMEE}{weighted minimum energy efficiency}
\newacronym{kkt}{KKT}{Karush-Kuhn-Tucker}
\newacronym{pc}{PC}{pseudo-concave}
\newacronym{qc}{QC}{quasi-concave}
\newacronym{ql}{QL}{quasi-linear}
\newacronym{pl}{PL}{pseudo-linear}
\newacronym{spc}{SPC}{strictly pseudo-concave}
\newacronym{sqc}{SQC}{strictly quasi-concave}
\newacronym{lfp}{LFP}{linear fractional problem}
\newacronym{clfp}{CLFP}{concave-linear fractional problem}
\newacronym{ccfp}{CCFP}{concave-convex fractional problem}
\newacronym{mmfp}{MMFP}{max-min fractional problem}
\newacronym{sorp}{SoRP}{sum-of-ratios problem}
\newacronym{porp}{PoRP}{product-of-ratios problem}
\newacronym{srp}{SRP}{single-ratio problem}
\newacronym{brb}{BRB}{branch-reduce-and-bound}
\newacronym{qos}{QoS}{quality-of-service}
\newacronym{comp}{CoMP}{cooperative multi-point}
\newacronym{ue}{UE}{user equipment}
\newacronym{bs}{BS}{base station}
\newacronym{mrc}{MRC}{maximum ratio combining}
\newacronym{d2d}{D2D}{device-to-device}
\newacronym{lmmse}{LMMSE}{linear minimum mean square error}
\newacronym{ris}{RIS}{reconfigurable intelligent surface}

\usepackage{accents}

\usepackage[centerlast,small]{caption}

\begin{document}
\title{\huge On The Optimal Number of Reflecting Elements for Reconfigurable Intelligent Surfaces}

\author{Alessio Zappone,~\IEEEmembership{Senior Member~IEEE}, Marco Di Renzo,~\IEEEmembership{Fellow~IEEE}, Xiaojun Xi,~\IEEEmembership{Student Member,~IEEE}, Merouane Debbah,~\IEEEmembership{Fellow~IEEE}
	\thanks{A. Zappone is with the University of Cassino and Southern Lazio, Cassino, Italy (alessio.zappone@unicas.it). M. Di Renzo and X. Xi are with the Universit\'e Paris-Saclay, CNRS and CentraleSup\'elec, Laboratoire des Signaux et Syst\`emes,  91192 Gif-sur-Yvette, France (marco.direnzo@centralesupelec.fr), M. Debbah is with Huawei France R\&D, Boulogne-Billancourt, France (merouane.debbah@huawei.com).}}

\maketitle

\begin{abstract}
This work considers a point-to-point link where a reconfigurable intelligent surface assists the communication between  transmitter and receiver. The system rate, energy efficiency, and their trade-off are optimized with respect to the number of individually tunable elements of the intelligent surface. The resource allocation accounts for the communication phase and for the overhead due to channel estimation and to reporting the optimized resource allocation to the  intelligent surface. Numerical results confirm the optimality of the proposed methods and show the potential gains of reconfigurable intelligent surfaces. 
\end{abstract}

\section{Introduction} \label{Introduction}
Recently, the concept of smart radio environment has emerged as a candidate architecture for future 6G networks   \cite{RIS_GE_JSAC, TCOM_AI,SmartWireless,HuangHolo2020,BasarRIS,Ntontin2019}, wherein \glspl{ris} are used to coat environmental objects that are present in the propagation environment. An \gls{ris} is a planar structure made of several individually tunable elements, called meta-atoms or passive scatterers, that can be programmed and appropriately reconfigured to control the phase of the incoming electromagnetic signal, by reflecting or refracting it towards specified locations \cite{RIS_GE_JSAC}. In \cite{Qian2020} and \cite{Zhou2020b} the \gls{snr} scaling law of \gls{ris}-aided transmission is characterized, and the impact of hardware impairments are analyzed, respectively. 
Experimental testbeds have confirmed the potential gains of \glspl{ris} \cite{Tang2019,Southest_1,Southest_2,Dai2019}.

In the context of radio resource allocation, several works have appeared. In \cite{EE_RISs}, the rate and \gls{ee} of an \gls{ris}-based multiple input single output (MISO) downlink system are optimized by means of alternating optimization, fractional programming, and sequential optimization. A similar system setup is addressed in \cite{Wu2018}, and the problem of power minimization subject to minimum rate constraints is tackled by alternating optimization. In \cite{Yang2019} and \cite{Yu2019}, sum-rate maximization in a MISO downlink system with orthogonal frequency division multiplexing is studied by optimizing the transmit beamformer and the \gls{ris} phase shifts with the aid of  alternating optimization. An \gls{ris} with discrete phase shifts is considered in \cite{Liu2019}, where sum-rate maximization in a multi-user MISO system is addressed. In \cite{Jiang2019}, a multi-user MISO channel is considered, in which an \gls{ris} is used to perform over-the-air computations. Alternating optimization and difference convex programming are used for system optimization. In \cite{Li2019b}, multiple \glspl{ris} are used in a massive \gls{mimo} setup, and the problem of maximizing the minimum of the users' signal-to-interference-plus-noise-ratio with respect to the transmit precoder and the \glspl{ris} phase shifts is considered. In \cite{Han2019b}, the problem of power control for physical-layer broadcasting under quality of service constraints for the mobile users is addressed. The downlink of a MIMO multi-cell system is considered in \cite{Pan2019b}, wherein the problem of weighted sum-rate maximization is tackled by alternatively optimizing the base station beamformer and the \gls{ris} phase shifts. An \gls{ris}-based MISO millimiter-wave system is studied in \cite{Wang2019b}, wherein the transmit beamforming and the phase shifts of multiple \glspl{ris} are optimized. In \cite{You2019}, channel estimation and sum-rate maximization are performed for a single-user uplink \gls{ris}-based system, by considering an \gls{ris} with discrete phase shifts. In \cite{PanJSAC2020}, the sum-rate of a MIMO RIS-based system employing simultaneous information and power transfer is maximized with respect to the transmitter beamforming and the RIS phase shifts. Power control for physical layer broadcasting in an \gls{ris}-based network is investigated in \cite{Han2020}. Rate maximization for \gls{ris}-based indoor millimeter-waves communications is addressed in \cite{Perovic2020} by adjusting the \gls{ris} and the transmitter phase shifts.

The above literature survey shows that most works focus on how to maximize the rate of \gls{ris}-based systems with respect to the transmit power/beamforming and to the \gls{ris} phase shifts, whereas the number of tunable elements at the \gls{ris} is not optimized. Moreover, previous works consider only the data transmission phase, while neglecting the overhead for channel estimation and reporting the optimized configuration to the \gls{ris}. In contrast to these research works, this paper addresses the problem of jointly optimizing the \gls{ris} phase shifts and the number of tunable elements to be activated at the \gls{ris}, in order to optimize the rate, the \gls{ee} and their trade-off. Notably, this is performed based on the model recently proposed in \cite{ZapponeTWC2020}, which quantifies the impact of channel estimation and resource allocation feedback on the rate and \gls{ee} of \gls{ris}-based systems.  

The rest of the paper is organized as follows. Section \ref{Sec:Model} describes the system model, Sections \ref{Sec:OptRateN}, \ref{Sec:EEOptimization}, and \ref{Sec:BiObjectiveOpt} develop the proposed algotihm for the optimization of the rate, \gls{ee}, and their trade-off, respectively. Section \ref{Sec:Numerics} numerically analyzes the proposed algorithms, while concluding remarks are given in Section \ref{Sec:Conclusions}.

\section{System Model and Problem Statement}\label{Sec:Model}
We consider a point-to-point system wherein a single-antenna transmitter and receiver communicate through an RIS equipped with $N$ individually tunable elements. This  scenario models, for example, a device-to-device communication link, or a cellular network in which the base stations employ antenna selection to serve each user by a different antenna, and multi-user interference is suppressed by orthogonal frequency division multiple access or any other orthogonal signaling technique. Also, numerical results in \cite{ZapponeTWC2020} have  shown that RISs are especially useful when few antennas are employed, since a proper RIS design can compensate for the lack of multiple transmit and/or receive antennas, without suffering a large overhead for channel estimation and feedback.

We assume that the direct path between transmitter and receiver is not available, and denote by $\bh=\{h_{n}\}_{n=1}^{N}$ and $\bg=\{g_{n}\}_{n=1}^{N}$ the fast-fading vector channels from the transmitter to the \gls{ris} and from the \gls{ris} to the receiver, and by $\delta$ the overall propagation path-loss. Before data communication can take place, the channels $\bh$ and $\bg$ must be estimated and the \gls{ris} configuration must be optimized and deployed at the \gls{ris}. Channel estimation and \gls{ris} optimization can take place at either the transmitter or the receiver through traditional pilot signaling techniques, while the optimized \gls{ris} configuration is implemented thanks to an \gls{ris} controller with minimal signal processing, transmission/reception, and power storage capabilities. It should be stressed that the presence of the controller is essential in order for the \gls{ris} to be dynamically reconfigurable in response to changes of the propagation channels \cite[Fig. 4]{RIS_GE_JSAC}. On the other hand, sending the control signal to the \gls{ris} introduces a non-negligible overhead to the communication phase, especially for large $N$. Denoting by $T$ the total duration of the time slot comprising the channel estimation phase of duration $T_E$, the  control feedback phase of duration $T_F$, and the data communication phase of duration $T-T_{E}-T_{F}$, and defining $\beta=p\delta/(BN_{0})$, with $B$ the communication bandwidth, $N_{0}$ the receive power spectral density, and $p$ the transmit power, the system rate and \gls{ee} are expressed as
\begin{align}\label{Eq:BitsTx}
R(\!N\!)&=\left(1-\frac{T_{E}+T_{F}}{T}\right)B\log\left(1+\beta |\bg^{H}\bPhi\bh|^{2}\right)\\
\text{EE}(N)&=R(N)/P_{tot}\label{Eq:EE}
\end{align}
wherein $\bPhi=\text{diag}(e^{j\phi_{1}},\ldots,e^{j\phi_{N}})$ is the \gls{ris} phase matrix and $P_{tot}$ is the total power consumption in the whole timeframe $T$. Following the model developed in \cite{ZapponeTWC2020}, we have $T_{E}=T_{0}(N+1)$, with $T_{0}$ the duration of each pilot tone, and
\begin{align} 
\small
T_{F}&\!=\!\frac{Nb_{F}}{B_{F}\log(1+p_{F}|h_{F}|^{2}/(N_{0}B_{F}))}\\
P_{tot}&\!=\!P_{E}\!+\!(1\!-\!T_{E}/T)\mu p\!+\!T_{F}/T(\mu_{F}p_{F}\!-\!\mu p)\!+\!NP_{c,n}\!+\!P_{c,0},\notag
\end{align}
since a power $p$ is used for $T-T_{E}-T_{F}$ seconds, with transmit amplifier efficiency $1/\mu$, a power $p_{F}$ is used for $T_{F}$ seconds, with transmit amplifier efficiency $1/\mu_{F}$, while hardware static power is consumed during the whole interval $T$, where $P_{c,n}$ is the hardware power required for each \gls{ris} element, $P_{c,0}$ is the hardware power for all other system components, and $P_{E}=T_{E}P_{0}/T$ is the power consumption during the channel estimation phase, with $P_{0}$ the power of each pilot tone. 

The aim of this work is to optimize the RIS phase shift matrix $\bPhi$ and the number of tunable elements $N$ at the RIS, in order to optimize the system rate and \gls{ee} in \eqref{Eq:BitsTx}, and their trade-off. Note that the approach in this work differs from robust resource allocation methods which assume imperfect channel state information \cite{Zhou2020,Hong2020,Zhou2020c}. Indeed, we assume that reliable channel estimation is performed and the resulting overhead is appropriately accounted for in our system model.

\subsection{Optimization of $\bPhi$ And Upper-Bound of $N$}\label{Sec:PhaseOpt}
Since $\bPhi$ does not affect the power consumption $P_{tot}$, the optimal $\bPhi$ for both the rate and the \gls{ee} is obtained by maximizing the term $|\bg^{H}\bPhi\bh|^{2}$. It is easy to see that this is accomplished by setting $\phi_{n}=\angle{g_{n}^{*}h_{n}}$ for any $n$. With this choice, the received power at the destination is $p\delta\left(\sum_{n=1}^{N}\alpha_{n}\right)^{2}$, with $\alpha_{n}=|h_{n}g_{n}|$. On the other hand, since the received power can not be larger than the transmit power, it must hold that $\delta\left(\sum_{n=1}^{N}\alpha_{n}\right)^{2}\leq 1$. Defining $\alpha_{max}=\max_{n}\alpha_{n}$, a sufficient condition for this to hold is $N\alpha_{max}\sqrt{\delta}\leq 1$, which sets an upper-bound on the maximum number of elements that can be placed on the RIS in order for the considered path-loss model to be physically meaningful. In addition, in order to prove the mathematical results to follow, we assume $\beta\alpha_{max}^{2}\geq 1$. This appears realistic, since $\beta\alpha_{max}^{2}$ is the receive SNR over the reflection path with the largest gain. Moreover, since $N\alpha_{max}\sqrt{\delta}\leq 1$, in order to enforce $\beta\alpha_{max}^{2}\geq 1$ we must have $N^{2}\leq \beta/\delta$, which finally allows us to bound the maximum number $N_{max}$ of elements at the \gls{ris} as 
\beq
N_{max} \leq\min\left\{(\alpha_{max}\sqrt{\delta})^{-1},\sqrt{p\beta/\delta}\right\}.
\eeq

\section{Rate optimization}\label{Sec:OptRateN}
Plugging the optimal $\bPhi$, the rate maximization problem is
\begin{align}\label{Eq:OptN}
&\max_{1\leq N\leq \min\{N_{max},\lfloor c/d\rfloor\}}  \!(c\!-\!dN)B\log\!\left(\!1\!+\!\beta\!\left(\!\sum_{n=1}^{N}\alpha_{n}\!\right)^{2}\!\right)
\end{align}
with $c=1-\frac{T_{0}}{T}$, $d=\frac{T_{0}}{T}+\frac{b_{F}}{TB_{F}\log\left(1+\frac{p_{F}|h_{F}|^{2}}{B_{F}N_{0}}\right)}$, and where, without loss of generality, we consider that the coefficients $\alpha_{n}$ are arranged in decreasing order of magnitude, i.e. $\alpha_{n}\geq  \alpha_{n+1}$ for all $n=1,\ldots,\min\left\{N_{max}, \left\lfloor\frac{c}{d}\right\rfloor\right\}$, which also means that $\alpha_{max}=\alpha_{1}$. The constraint in \eqref{Eq:OptN} ensures that the sum of the  durations of the channel estimation and feedback phases is shorter than the total length of the frame, and that $N$ is smaller than its maximum feasible number $N_{max}$. The challenge in solving \eqref{Eq:OptN} lies in the fact that the first factor of the objective decreases with $N$, while the second factor increases with $N$, which makes it difficult to determine the behavior of the rate $R$ as a function of the discrete variable $N$. In order to globally solve \eqref{Eq:OptN}, the following result is instrumental.  
\begin{proposition}\label{Prop:Rate_N}
$R(N)$ in \eqref{Eq:OptN} is a unimodal function, i.e. it is either increasing with $N$ or, if there exists an $\bar{N}$ such that $R(\bar{N})\geq R(\bar{N}+1)$, $R(N)$ is decreasing for $N\geq \bar{N}$.
\end{proposition}
\begin{IEEEproof}
	If $\bar{N}$ does not exist, the rate is increasing with $N$. If $\bar{N}$ exists, then defining  $f(N)\!=\!B\log\!\left(\!1\!+\!\beta\!\left(\sum_{n=1}^{N}\alpha_{n}\!\right)^{2}\!\right)$, the 
condition $R(\bar{N})\geq R(\bar{N}+1)$ is equivalent to
\beq\label{Eq:RateConditionN}
(c-d\bar{N})(f(\bar{N}+1)-f(\bar{N}))\leq d\; f(\bar{N}+1)\;.
\eeq
Thus, the result holds if we can show that \eqref{Eq:RateConditionN} implies that $R(\bar{N}+1)\geq R(\bar{N}+2)$, i.e.
\beq\label{Eq:RateConditionN1}
(c-d(\bar{N}+1))(f(\bar{N}+2)-f(\bar{N}+1))\leq d\; f(\bar{N}+2)\;.
\eeq
At this point, let us show that, for any $N$, it holds
\beq\label{Eq:DecIncrement}
f(N+1)-f(N)\geq f(N+2)-f(N+1)\;.
\eeq
To see this, expanding the square in $f(N+1)$ leads to
\begin{align}
&f(N+1)-f(N)=\\
&B\log\!\!\left(\!1\!+\!\frac{\beta\left(\!\!\left(\sum_{n=1}^{N}\alpha_{n}\right)^{2}\!\!\!+\!\alpha_{N+1}^{2}\!\!+\!2\alpha_{N+1}\sum_{n=1}^{N}\alpha_{n}\!\right)}{1+\beta\left(\sum_{n=1}^{N}\alpha_{n}\right)^{2}}\!\right)\!\!=\notag\\
&B\!\log\!\!\left(\!\!1+\frac{\beta\alpha_{N+1}\left(\alpha_{N+1}+2\sum_{n=1}^{N}\alpha_{n}\right)}{1+\beta\left(\sum_{n=1}^{N}\alpha_{n}\right)^{2}}\!\!\right)\notag
\end{align}
Similarly, it holds
\begin{equation}
f(N+2)\!-\!f(N+1)\!=\!B\!\log\!\!\left(\!\!1+\!\frac{\beta\alpha_{N+2}\!\left(\!\alpha_{N+2}\!+\!2\sum_{n=1}^{N+1}\alpha_{n}\right)}{1\!+\!\beta\left(\sum_{n=1}^{N+1}\alpha_{n}\!\right)^{2}}\!\right).\notag
\end{equation}
Then, the condition in \eqref{Eq:DecIncrement} becomes
\begin{align}\label{Eq:DecIncrement2}
\frac{\alpha_{N+1}^{2}}{1+\beta\left(\sum_{n=1}^{N}\alpha_{n}\right)^2}+\frac{2\alpha_{N+1}\sum_{n=1}^{N}\alpha_{n}}{1+\beta\left(\sum_{n=1}^{N}\alpha_{n}\right)^{2}}\geq \notag\\
\frac{\alpha_{N+2}^{2}}{1+\beta\left(\sum_{n=1}^{N+1}\alpha_{n}\right)^2}+\frac{2\alpha_{N+2}\sum_{n=1}^{N+1}\alpha_{n}}{1+\beta\left(\sum_{n=1}^{N+1}\alpha_{n}\right)^{2}}\;.
\end{align}
Since $\alpha_{n}\geq\alpha_{n+1}$ for any $n=1,\ldots,N$, the first summand at the left-hand-side of \eqref{Eq:DecIncrement2} is larger than the first summand at the right-hand-side. Then, a sufficient condition for \eqref{Eq:DecIncrement2} to hold is
\begin{align}\label{Eq:DecIncrement3}
\frac{\sum_{n=1}^{N}\alpha_{n}}{1+\beta\left(\sum_{n=1}^{N}\alpha_{n}\right)^{2}}\geq
\frac{\sum_{n=1}^{N+1}\alpha_{n}}{1+\beta\left(\sum_{n=1}^{N+1}\alpha_{n}\right)^{2}}\;.
\end{align}
Defining $z=\sum_{n=1}^{N}\alpha_{n}$, it can be seen that \eqref{Eq:DecIncrement3} is equivalent to showing that the function $g(z)=\frac{z}{1+\beta z^{2}}$ is decreasing. Computing the first-order derivative of $g(z)$, and setting it to be non-positive, yields $\beta\alpha_{1}\geq 1$, which holds on the feasible set of \eqref{Eq:OptN}. Finally, exploiting \eqref{Eq:RateConditionN}, and observing that $f(N)$ is increasing in $N$ since each $\alpha_{n}$ is positive, it follows that
\begin{align}
&(c-d(\bar{N}+1))(f(\bar{N}+2)-f(\bar{N}+1))\leq\\
&(c-d\bar{N})(f(\bar{N}+1)-f(\bar{N}))\leq\notag\\
& d\;f(\bar{N}+1)\leq d\;f(\bar{N}+2)\;,\notag
\end{align}
and hence the proof follows.
\end{IEEEproof}
Equipped with this result, Problem \eqref{Eq:OptN} can be solved by a greedy approach in which the tunable elements of the \gls{ris} are activated one at a time, following the decreasing order of magnitude of the coefficients $\{\alpha_{n}\}_n$, until a decrease in $R(N)$ is observed, or $N$ reaches its maximum allowed number. The procedure is stated in Algorithm \ref{Alg:OptN}, with $A(N)=R(N)$.
\begin{algorithm}\caption{Optimization of $N$ for rate maximization}
\begin{algorithmic}\label{Alg:OptN}
\STATE $N=1$; $A_{0}=0$; $A_{1}=A(N)$;
\WHILE{$A_{n}\geq A_{n-1}$ \texttt{and} $N\leq \min\left\{N_{max},\left\lfloor\frac{c}{d}\right\rfloor\right\}$}
\STATE $N=N+1$; 
\STATE $A_{n}=A(N)$;
\ENDWHILE
\end{algorithmic}
\end{algorithm}

\section{Energy efficiency optimization}\label{Sec:EEOptimization}
Defining $\gamma=P_{c,0}+P_{0}\frac{T_{0}}{T}+\mu pc$ and $\psi=(P_{0}-\mu p-1)\frac{T_{0}}{T}+d+P_{c,n}$, the  \gls{ee} maximization problem is stated as 
\begin{subequations}
\begin{align}\label{Eq:EE_N}
&\ds\max_{N}\frac{(c\!-\!d N)f(N)}{\gamma+\psi N}\\
&\;\textrm{s.t.}\;1\!\leq\! N\!\leq \!\min\left\{\!N_{max}, \lfloor c/d\rfloor\!\right\}\;.
\end{align}
\end{subequations}
Proposition \ref{Prop:EE_N} shows that the \gls{ee} is unimodal, and thus \eqref{Eq:EE_N} can be globally solved by Algorithm \ref{Alg:OptN} with $A(N)=\text{EE}(N)$. 
\begin{proposition}\label{Prop:EE_N}
$\text{EE}(N)$ in \eqref{Eq:EE_N} is a unimodal function, i.e. it is either increasing with $N$ or, if there exists an $\bar{N}$ such that $\text{EE}(\bar{N})\geq \text{EE}(\bar{N}+1)$, $\text{EE}(N)$ is decreasing for $N\geq \bar{N}$.
\end{proposition}
\begin{IEEEproof}
	Proceeding like in the proof of Proposition \ref{Prop:Rate_N}, if $\bar{N}$ does not exist, then $\text{EE}(N)$ is increasing with $N$. Instead, if $\bar{N}$ exists, the result follows if we can prove that $\text{EE}(\bar{N)}\geq \text{EE}(\bar{N}+1)$ implies that $\text{EE}(\bar{N}+1)\geq \text{EE}(\bar{N}+2)$. First of all, let us observe that if $\bar{N}$ is such that $R(\bar{N})\geq R(\bar{N}+1)$, then we already know from Proposition \ref{Prop:Rate_N} that $\bar{N}$ falls in the range in which the rate function $R(N)$ is decreasing. This implies that the \gls{ee} is decreasing for any $N\geq \bar{N}$, since increasing $N$ yields a lower numerator and a larger denominator. As a result, the non-trivial case to be considered is when $\bar{N}$ belongs to the range in which the rate is still increasing, i.e. the first decrease in the \gls{ee} happens when the rate function is still increasing with $N$. Thus, without loss of generality, in the rest of this proof we assume $R(\bar{N}+2)>R(\bar{N}+1)>R(\bar{N})$. Next, let us observe that $\text{EE}(\bar{N)}\geq \text{EE}(\bar{N}+1)$ is equivalent to
    \beq\label{Eq:EE_N1}
    \bar{N}\! \leq\! -\frac{\gamma}{\psi}+\frac{R(\bar{N})}{R(\bar{N}\!+\!1)\!-\!R(\bar{N})}\;. 
    \eeq
    Similarly, $\text{EE}(\bar{N}+1)\geq \text{EE}(\bar{N}+2)$ is equivalent to
    \beq\label{Eq:EE_N2}
    \bar{N}\! \leq\! -\frac{\gamma}{\psi}-1+\frac{R(\bar{N}+1)}{R(\bar{N}\!+\!2)\!-\!R(\bar{N}+1)}\;.
    \eeq
    At this point, we note that \eqref{Eq:EE_N1} can be written as
    \beq
    \bar{N}\! \leq\!-\frac{\gamma}{\psi}-1+\frac{R(\bar{N}\!+\!1)}{R(\bar{N}\!+\!1)\!-\!R(\bar{N})}\;,
	\eeq
	which implies \eqref{Eq:EE_N2} if we can show that $R(\bar{N}+2)-R(\bar{N}+1)\leq R(\bar{N}+1)-R(\bar{N})$. To show this, we observe that we have
	\begin{align}
	&R(\bar{N}+2)-R(\bar{N}+1)\\
	&\hspace{0.8cm}=\left(c-d(\bar{N}+1)\right)\left(f(\bar{N}+2)-f(\bar{N}+1)\right)-df(\bar{N}+2)\notag\\
	&\hspace{0.8cm}\leq\left(c-d(\bar{N}+1)\right)\left(f(\bar{N}+2)-f(\bar{N}+1)\right)-df(\bar{N})\notag\\	&\hspace{0.8cm}\leq\left(\!c\!-\!d(\bar{N}\!+\!1)\!\right)\!\left(f(\bar{N}\!+\!1)\!-\!f(\bar{N})\!\right)\!-\!df(\bar{N})\notag\\
	&\hspace{0.8cm}=R(\bar{N}\!+\!1)\!-\!R(\bar{N})\notag\;,
	\end{align}
	where the two equalities follow recalling that $R(N)=(c-dN)f(N)$, while the two inequalities follow since $f(N)$ is non-negative, increasing, and such that $f(N+1)-f(N)\geq f(N+2)-f(N+1)$, as proved in Proposition \ref{Prop:Rate_N}. 
\end{IEEEproof}

\section{Rate-EE maximization}\label{Sec:BiObjectiveOpt}
Rate-EE maximization is cast as the bi-objective problem
\begin{align}\label{Eq:Trade-off}
&\max_{N}\big\{R(N),\text{EE}(N)\big\}\;,\;\textrm{s.t.}\;1\!\leq \!N\!\leq\! \min\left\{\!N_{max}, \lfloor c/d\rfloor\!\right\}
\end{align}
By virtue of \cite[Th. 3.4.5]{MOBook}, all Pareto-optimal points of \eqref{Eq:Trade-off} can be obtained by solving
\begin{subequations}
\begin{align}\label{Eq:Trade-offMaxMin0}
&\max_{N}\min\{w(R(\!N\!)\!-\!R_{opt}),\!(1\!-\!w)(\text{EE}(\!N\!)\!-\!\text{EE}_{opt})\}\\
&\;\textrm{s.t.}\; 1\leq N\leq\min\left\{\!N_{max}, \left\lfloor\frac{c}{d}\right\rfloor\!\right\}
\end{align}
\end{subequations}
for $w\in(0,1)$, with $R_{opt}$ and $\text{EE}_{opt}$ the individual maximizers of $R(N)$ and $\text{EE}(N)$, that as derived in Sections \ref{Sec:OptRateN} and \ref{Sec:EEOptimization}. The following proposition shows that $G(N)=\min\{w(R(\!N\!)\!-\!R_{opt}),\!(1\!-\!w)(\text{EE}(\!N\!)\!-\!\text{EE}_{opt})\}$ is unimodal, and thus  Problem \eqref{Eq:Trade-offMaxMin0} can be solved by Algorithm \ref{Alg:OptN} with $A(N)=G(N)$. 
\begin{proposition}\label{Prop:TradeOff_N}
$G(N)$ is a unimodal function, i.e. it is either increasing with $N$, or, if there exists $\bar{N}$ such that $G(\bar{N})\geq G(\bar{N}+1)$, $G(N)$ is decreasing for $N\geq \bar{N}$.
\end{proposition}
\begin{IEEEproof}
Propositions \ref{Prop:Rate_N} and \ref{Prop:EE_N} ensure that $R(N)$ and $\text{EE}(N)$ are unimodal functions, which implies that $G_{1}=w(R(N)-R_{opt})$ and $G_{2}=(1-w)(\text{EE}(N)-\text{EE}_{opt})$ are unimodal, too. 

Let us consider first that both $G_{1}(N)$ and $G_{2}(N)$ have a finite maximizer, which we denote by $N_{1}$ and $N_{2}$, respectively. Without loss of generality, let us assume that $N_{1}\leq N_{2}$. Then,  $G(N)$ is increasing for $N\leq N_{1}$ and decreasing for $N\geq N_{2}$. As for the range $N_{1}<N<N_{2}$, let us consider two cases:

(a) If $G_{1}(N_{1})\leq G_{2}(N_{1})$, then, $G(N)=G_{1}(N)$ for $N_{1}<N<N_{2}$, because in this range $G_{2}$ is increasing while $G_{1}$ is decreasing. As a result, $G(N)$ is increasing for $N\leq N_{1}$ and decreasing for $N>N_{1}$ and we have $\bar{N}=N_{1}$.

(b) If $G_{1}(N_{1})>G_{2}(N_{1})$, two cases can be considered: if $G_{1}(N)\geq G_{2}(N)$ for $N_{1}\leq N\leq N_{2}$, then $G(N)=G_{2}(N)$ for $N_{1}\leq N\leq N_{2}$ and the thesis holds with $\bar{N}=N_{2}$; if instead  $G_{1}(N)\not\geq G_{2}(N)$ for $N_{1}\leq N\leq N_{2}$, then we can define $N_{3}$ as the smallest number such that $N_{1}< N_{3}\leq N_{2}$ and $G_{1}(N_{3})\leq G_{2}(N_{3})$. Thus, $G(N)=G_{2}(N)$ for $N_{1}\leq N< N_{3}$, while $G(N)=G_{1}(N)$ for $N_{3}\leq N\leq N_{2}$. Then, it follows that $G(N)$ is increasing for $N< N_{3}$ and decreasing for $N\geq N_{3}$, and the thesis follows with $\bar{N}=N_{3}$.

Finally, we observe that the reasoning above includes as special cases the situations in which either $G_{1}$ or $G_{2}$ is monotonically increasing for all $N$, while in the case in which both $G_{1}$ and $G_{2}$ are monotonically increasing $G(N)$ is monotonically increasing and $\bar{N}$ does not exist.
 \end{IEEEproof} 

\section{Numerical results}\label{Sec:Numerics}
In our numerical simulation we set $P_{F}=30\;\textrm{dBm}$, $B=5\,\textrm{MHz}$, $B_{F}=1\,\textrm{MHz}$, $\delta=110\,\textrm{dB}$, $N_{0}=-174\; \textrm{dBm/Hz}$, $\mu=\mu_{F}=1$, $P_{c,0}= 45\;\textrm{dBm}$, $P_{c,n}= 10\;\textrm{dBm}$, $b_{F}=16$, $T_{0}=0.5\,\textrm{ms}$, $P_{0}=10\,\textrm{dBm}$. Moreover, $h_{n}\sim{\cal CN}(v_{h},1)$ and $g_{n}\sim{\cal CN}(v_{g},1)$. Thus, $|h_{n}|$ and $|g_{n}|$ are Rice distributed, where $v_{h}$ and $v_{g}$ are chosen so that the power of the line-of-sight path is four times larger than the power of all other paths. A similar model is used for the feedback channel $h_{F}$. Also, $N_{max}=200$ and all presented results have been averaged over $10^{4}$ independent realizations of the channel vectors $\bh$ and $\bg$. 

\begin{figure}
\centering\includegraphics[width=\columnwidth]{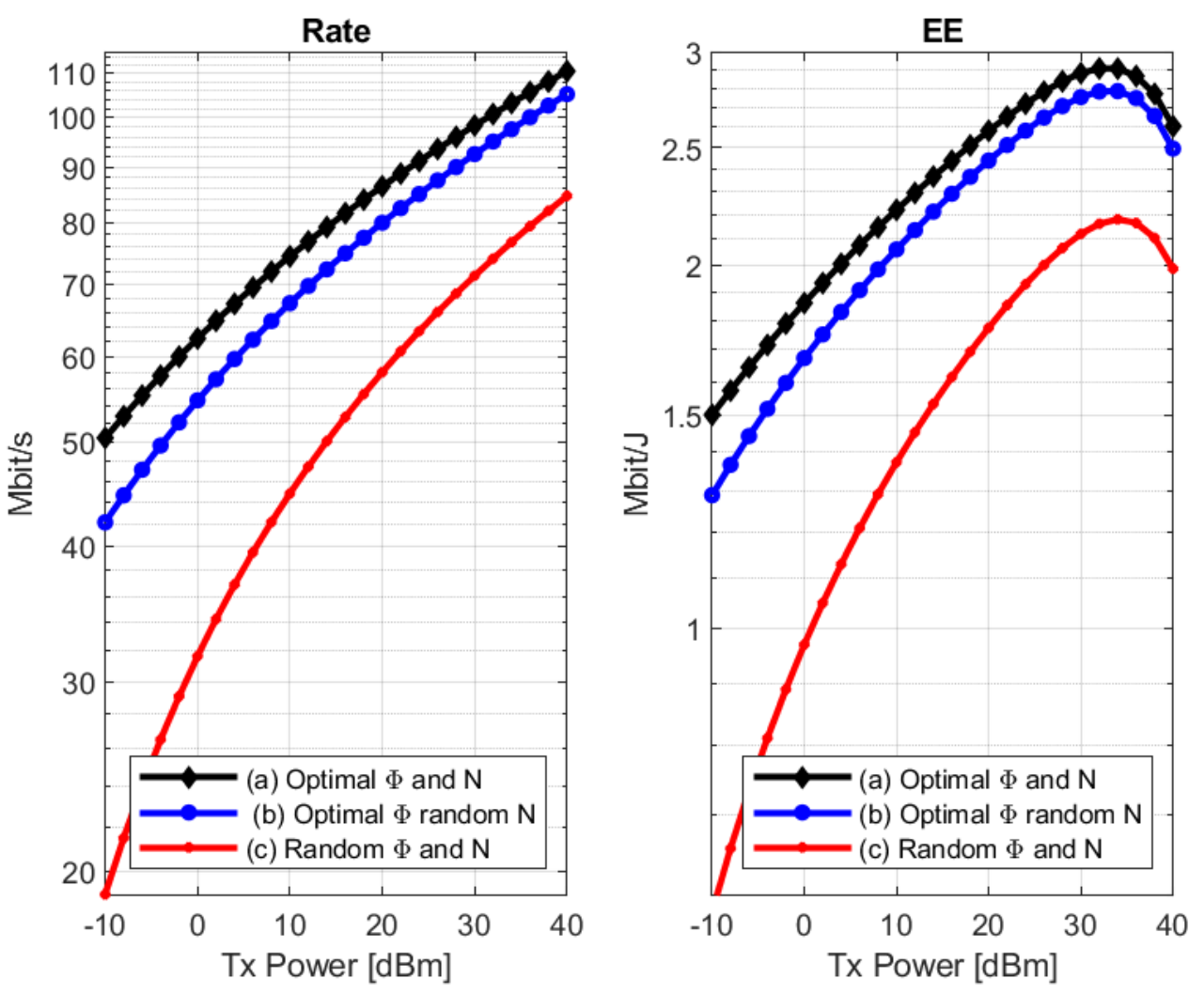}
	\caption{Rate (left) and EE (right) versus $P$ for: (a) Optimal $N$ and $\bPhi$; (b) Random $N$ and optimal $\bPhi$; (c) Random $N$ and $\bPhi$.}\label{Fig:Rate_EE}
	\end{figure}

Fig. \ref{Fig:Rate_EE} shows the rate and \gls{ee} versus the transmit power $P$ achieved by the following schemes:
\begin{itemize}
	\item [(a)] Optimization of $N$ by Algorithm \ref{Alg:OptN} and optimization of $\bPhi$ as described in Section \ref{Sec:PhaseOpt}.
	\item [(b)] Random selection of $N$ in the range $[1,\min\{N_{max},c/d\}]$ and optimization of $\bPhi$ as described in Section \ref{Sec:PhaseOpt}.
	\item [(c)] Random selection of $N$ in the range $[1,\min\{N_{max},c/d\}]$ and random selection of $\phi_{n}$ in $[0,2\pi]$ for all $n=1,\ldots,N$. Since no configuration of the \gls{ris} phase shifts is required for this scheme, we set $T_F=0$ and $T_E = T_0$ in this case.
\end{itemize}
The results show that accounting, at the design stage, for the overhead due to channel estimation and optimal \gls{ris} configuration outperforms random resource allocation. In particular, the proposed Scheme (a) grants a large gain compared to Scheme (c) which randomly selects both $N$ and $\bPhi$, despite the fact that no overhead is required in this case for reporting the resource allocation at the \gls{ris}.

Table \ref{Tab:Maximizer} reports the rate maximizer $N_{R}^{*}$ and the \gls{ee} maximizer $N_{\text{EE}}^{*}$ versus $P$ corresponding to the performance of Scheme (a) in Fig. \ref{Fig:Rate_EE} The results confirm that both the rate and the \gls{ee} have a finite maximizer with $N$.
\begin{table}
\centering
	\begin{tabular}{ | c | c | c | c | c | c |}
		\hline
		 & 0\,\textrm{dBm} & 10\,\textrm{dBm} & 20\,\textrm{dBm} & 30\,\textrm{dBm} & 40\,\textrm{dBm} \\
		\hline
		$N_{R}^{*}$ & 198.89 & 197.32 & 193.38 & 184.74 & 172.04 \\
		\hline
		$N_{\text{EE}}^{*}$ & 187.05 & 167.29 & 147.01 &  133.92 & 145.99 \\
		\hline
	\end{tabular}\caption{Network parameters}\label{Tab:Maximizer}
\end{table}

\begin{figure}[!h]
\centering	\includegraphics[width=\columnwidth]{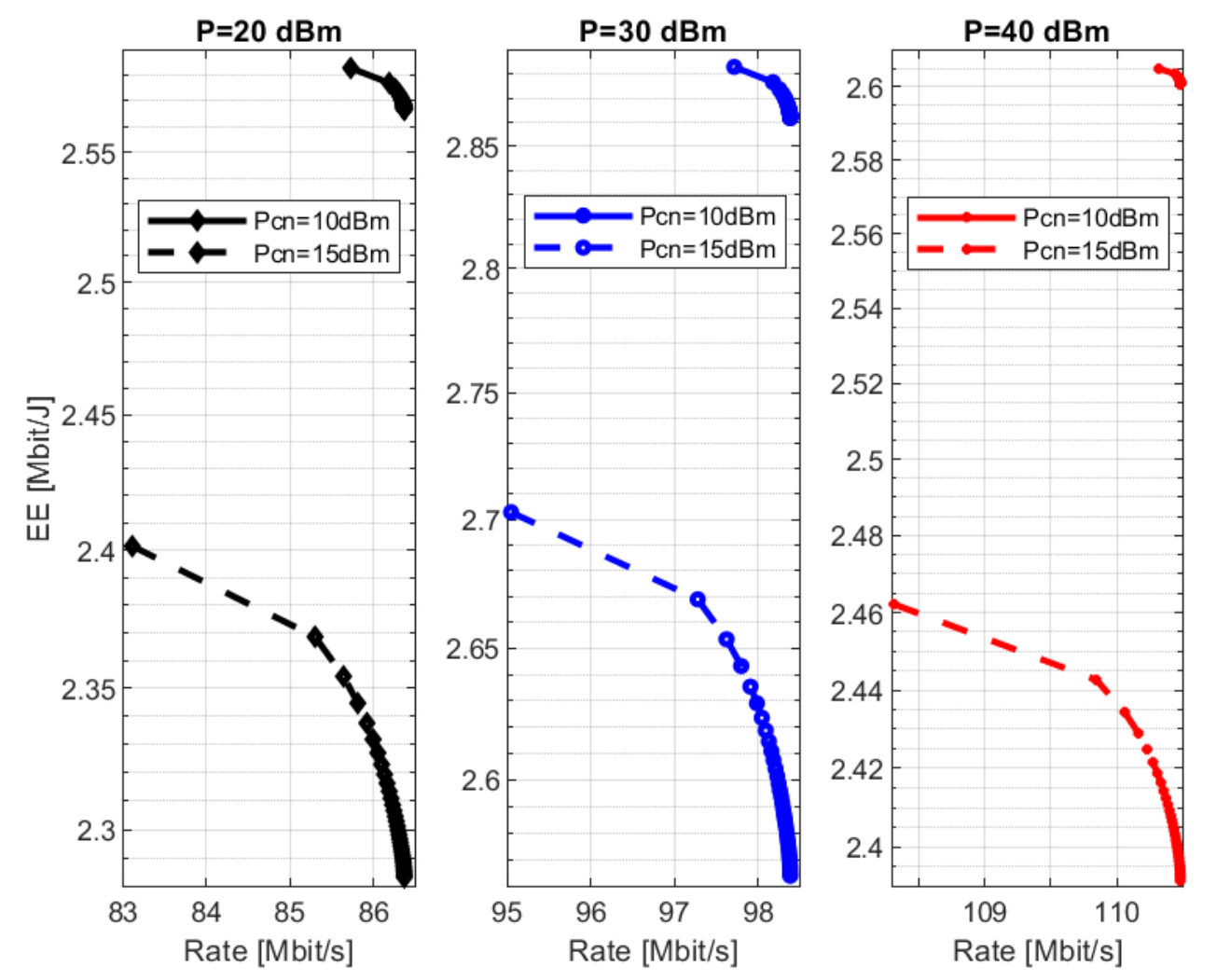}
	\caption{Rate-EE Pareto region for $P=20\,\textrm{dBm}$ (left), $30\,\textrm{dBm}$ (middle), $40\,\textrm{dBm}$ (right), with $P_{c,n}=10\,\textrm{dBm}$ and $15\,\textrm{dBm}$}
	\label{Fig:BiOpt}
\end{figure}

Finally Fig. \ref{Fig:BiOpt} shows the Rate-EE Pareto-frontier of the considered \gls{ris}-based system for $P=20\,\textrm{dBm}$, $P=30\,\textrm{dBm}$, $P=40\,\textrm{dBm}$. For each value of $P$ the cases $P_{c,n}=10\,\textrm{dBm}$ and $P_{c,n}=15\,\textrm{dBm}$ are shown. As expected, a higher $P_{c,n}$ yields a wider Pareto-region, because the higher $P_{c,n}$ is, the more the rate and the \gls{ee} are contrasting objectives. 

\section{Conclusions}\label{Sec:Conclusions}
This work has optimized the number of tunable elements to be activated on a \gls{ris}, by accounting at the design stage for the overhead due to channel estimation and \gls{ris} configuration. Globally optimal and low-complexity algorithms are derived for the optimization of the rate, the \gls{ee}, and their trade-off. Numerical results have confirmed  that overhead-aware resource allocation may significantly outperform sub-optimal methods, such as random \gls{ris} configuration.

\bibliographystyle{IEEEtran}
\bibliography{FracProg}

\end{document}